\documentclass[graybox]{svmult}

\usepackage{type1cm}         

\usepackage{makeidx}         
\usepackage{graphicx}        
\usepackage{multicol}        
\usepackage[bottom]{footmisc}

\usepackage{sectsty}

\usepackage{newtxtext}       %
\usepackage[varvw]{newtxmath}       

\usepackage{wasysym}
\usepackage{comment}

\usepackage[colorlinks,urlcolor=black,linkcolor=black,citecolor=black]{hyperref}
\usepackage{subcaption}

\usepackage{siunitx}
\RequirePackage{pgfplots}						
\RequirePackage{pgfplotstable}					
\RequirePackage{pgfmath}
\pgfplotsset{compat=newest}
\pgfplotsset{plot coordinates/math parser=true}
\usepgfplotslibrary{groupplots}
\RequirePackage{tikz}
\RequirePackage[outline]{contour}					
\RequirePackage{tikz-3dplot}                     
\usetikzlibrary{%
  arrows,%
  calc,%
  fit,%
  intersections,  through,  3d,  matrix,  patterns,%
  plotmarks,%
  shapes.geometric,%
  shapes.misc,%
  shapes.symbols,%
  shapes.arrows,%
  shapes.callouts,%
  shapes.multipart,%
  automata,%
  knots,backgrounds,%
  chains,%
   topaths,%
    trees,%
     petri,%
      mindmap,%
       matrix,%
        folding,%
          fadings,%
           through,%
            positioning,%
             scopes,%
              decorations.fractals,%
               decorations.shapes,%
                decorations.text,%
                 decorations.pathmorphing,%
                  decorations.pathreplacing,%
                   decorations.footprints,%
                    decorations.markings,%
                    shadows,%
                    hobby,  spy, 
                    pgfplots.fillbetween,
                    pgfplots.external,
                    pgfplots.groupplots,
                    pgfplots.patchplots}
\usepackage[backend=biber, 
sorting=none,
doi=true,
isbn=false,
date=year, 
bibwarn = true,
url=false,
eprint=false,
firstinits = true,
maxbibnames= 300,
minbibnames= 300,
maxcitenames=200,
mincitenames=200,
style=numeric,
citestyle=numeric
]{biblatex} 
\bibliography{SFB.bib}

\usepackage{siunitx}
\makeindex             
                       
\usepackage{indentfirst}      
\usepackage{float}

\usepackage{chngcntr}
\counterwithout{equation}{chapter} 

\usepackage{makecell}

\usepackage{titlesec}
\titleformat*{\subsection}{\bf\normalsize\itshape}                 

\usepackage{enumitem}

\usepackage{todonotes}

\begin{document}




\pagestyle{empty}

\title{DEM simulation of the powder application in powder bed fusion}
\titlerunning{DEM simulation of the powder application in powder bed fusion}
\author{Vasileios Angelidakis, Michael Blank, Eric J. R. Parteli, Sudeshna Roy, Daniel Schiochet Nasato, Hongyi Xiao, and Thorsten P\"oschel}
\authorrunning{V. Angelidakis et al.}
%
%

\abstract*{Powder-based additive manufacturing is a powerful tool for the production of parts exhibiting high geometric complexity.
It finds applications where fast prototyping is sought, as it does not require the printing of supports, thus producing near-finished parts with little post-processing needed.
The many variants of this manufacturing approach involve the spreading of powder in layers, which are sintered by a heat source to the desired shape, only to be covered by the next layer, sequentially.
Powder layers have inherently inhomogeneous characteristics, which can be the source of defects in the final part, since granular materials tend to form unstructured packings.
Hence, increasing the spatial homogeneity of powder layer characteristics, such as density and surface roughness, is key in achieving high quality.
Micro-mechanical numerical simulations can be used to explore the effect of the environment conditions, material parameters and process variables that affect the homogeneity of powder layers.
In this work, we simulate powder spreading processes using the Discrete Element Method (DEM).
We study the effects of size, shape, temperature and cohesion  of the powder particles on the mechanical characteristics of powder layers. In addition, alternative application mechanisms of powders are explored to investigate their influence on the produced layer characteristics, and explore their efficiency of being used in real applications.}

\counterwithout{section}{chapter}

\maketitle

\section{Introduction}
The packing behavior of powders is significantly influenced by various types of inter-particle attractive forces, including adhesion and non-bonded van der Waals forces \cite{yu1997modelling,gotzinger2003dispersive,gotzinger2004particle,castellanos2005relationship,li2006london,severson2009mechanical}. Alongside particle size and shape distributions, the inter-particle interactions, in particular frictional and adhesive forces, play a crucial role in determining the flow behavior and consequently the packing density of the powder layer. The impact of various types of attractive forces on the packing density of powders with different materials and particle size distributions remains largely unexplored and requires further investigation. Accurately comprehending these effects through experiments while considering specific particle size distributions and material properties poses significant challenges. To address these challenges, we employ Discrete Element Method (DEM) simulations to characterize the packing behavior of fine powders. We can demonstrate quantitative agreement with experimental results by incorporating the appropriate particle size distribution and using an adequate model of attractive particle interactions. Furthermore, our findings indicate that both adhesion, which is modeled using the Johnson–Kendall–Roberts (JKR) model \cite{johnson1971surface}, and van der Waals interactions are crucial factors that must be taken into account in DEM simulations.

Characterization of the packing structure is essential to achieve simulation-driven optimization of the layer quality. While local packing density is commonly employed to assess granular packing, relying solely on this parameter is inadequate for identifying structural defects in disordered packing \cite{richard2020predicting}. This is due to the fact that numerous packing arrangements can exhibit the same local density. Conversely, local structural anisotropy is a fundamental characteristic of non-crystalline packing and plays a significant role in crucial mechanical properties within disordered packings, such as jamming \cite{rieser2016divergence}, plasticity \cite{richard2020predicting}, and shear band formation \cite{xiao2020strain,harrington2018anisotropic,harrington2020stagnant}. We use local structural anisotropy as a measure to characterize the structures of the deposited particles. This approach is threshold-free and provides a meaningful distribution that accurately reflects the variations in the packing structure within a deposited powder layer. It holds particular relevance in discerning structural differences between homogeneous packing observed in non-cohesive powders and the heterogeneous packing tendencies prevalent in highly cohesive powders.

The heat transfer within the powder bed is significantly impacted by the quality of the final powder layer, including aspects such as the packing density and surface profile \cite{zhao2023multiscale}. It is worth noting that many DEM simulations commonly disregard the influence of temperature on inter-particle interaction forces. Nevertheless, it is widely recognized that temperature can have a substantial impact on the mechanical behavior of particle interactions. The temperature dependence of viscosity is one of the most important variables for polymers, which is given by an Arrhenius type equation \cite{glasstone1941theory,jagota1993viscosities}. As the temperature approaches higher values, nearing the melting point, the particles within a stressed state experience a reduction in stiffness. Consequently, this leads to increased deformation and larger overlaps between particles due to the compressive forces acting upon them. Luding et al. \cite{luding2005discrete} introduced a new discrete model for the sintering of particulate materials with a temperature-dependent elastic modulus. 

The role of particle shape coupled with its thermal properties is hardly ever explored in the literature and thus is a subject of great potential. We propose an extension of the Discrete Element Method (DEM) that incorporates heat transfer capabilities by combining a multisphere algorithm with a thermal discrete particle model, enabling the simulation of non-spherical particles. To showcase the applicability of this model, we conduct simulations of a powder spreading process utilizing irregularly shaped Polyamide 12 (PA12) powder particles. This demonstration illustrates the effectiveness of the proposed model in capturing the behavior of non-spherical particles in real-world scenarios.

Improving the quality of the layer during powder spreading can be achieved by various means. In addition to exploring the thermal, cohesive, and overall mechanical properties of the powder material itself, an optimization of the steps of the powder application process can result in improved layer characteristics. To this end, the efficiency of a vibrating recoating mechanism has been explored with respect to the layer density and surface roughness.


\section{Particle-based tool for powder spreading}
\begin{figure}[htb!]
    \centering
    \begin{minipage}{0.5\textwidth}
        \begin{picture}(100,20)
            \put(0,0){\includegraphics[width=0.99\textwidth]{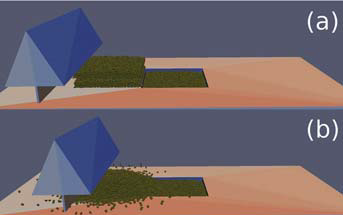}}
            \end{picture}
            \end{minipage}\hfill
                \begin{minipage}{0.5\textwidth}
        \begin{picture}(100,20)
            \put(0,0){\includegraphics[width=0.99\textwidth]{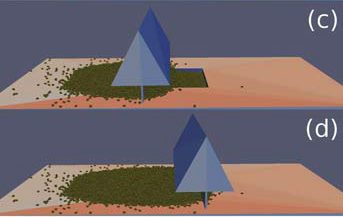}}
            \end{picture}
            \end{minipage}\hfill
\caption{Snapshots of a simulation of particles with complex geometric shapes. Particles constructed with the multisphere method are inserted into the system with dynamic boundary conditions that mimic the device used in additive manufacturing.}\label{fig:mimicpowderspreading}
\end{figure}

The first particle-based numerical tool was developed by \citeauthor{parteli2013simulation} \cite{parteli2013simulation} based on an existing DEM solver \cite{kloss2012models} to model the dynamics of geometrically complex particles subjected to dynamic boundary conditions. Snapshots of the simulation setup with complex shaped particles and dynamic boundary conditions are shown in \autoref{fig:mimicpowderspreading}. 
This setup mimics the device used for powder spreading in additive manufacturing. In this numerical setup, the moving boundaries (walls) are modeled by triangular meshes, which can be imported into the solver. The device consists of a rake for powder application (which moves from left to right in \autoref{fig:mimicpowderspreading}) and a building tank (central area) placed on top of a vertically adjustable platform. The simulation starts by releasing about $6500$ particles from a short distance above the platform, as shown in \autoref{fig:mimicpowderspreading}(a). After falling under gravity, the settled particles are transported on the platform as the rake moves from left to right, as shown in \autoref{fig:mimicpowderspreading}(b-d). This small setup helps us to address several questions relevant to the real powder spreading process, such as the role of particle shape and size distribution on the flowability of the powder within the device. This, in turn, allows us to understand the surface profile and packing density of the powder bed.

In \citeyear{parteli2016particle}, \citeauthor{parteli2016particle} \cite{parteli2016particle} established a real scale setup for the first time to simulate powder spreading using a roller as the spreading device, as shown in \autoref{fig:powderspreading}(a). The numerical tool based on the Discrete Element Method (DEM) accounts for a realistic description of inter-particle forces, particle size distributions, and complex geometric shapes of the powder particles, which play a major role in the static and dynamic characteristics of granular systems. With this numerical setup, the authors predicted the powder packing and surface roughness profile dependent on the recoating velocity and the powder particle size distribution. 

\section{Geometrically complex particles}
The first step to achieve reliable numerical simulations of the powder spreading application in additive manufacturing is the accurate representation of complex geometric shapes of the particles. While preliminary studies are done by \citeauthor{parteli2013using} \cite{parteli2013using,parteli2013simulation}, this representation is first accomplished by \citeauthor{parteli2016particle} \cite{parteli2016particle} in the context of powder spreading for powder bed fusion using the multisphere method, which consists of building clumps of spherical particles to model the complex shape of the target. Within the Discrete Element Method (DEM), there are various simulation techniques available for modeling non-spherical particles. One of the oldest and most versatile techniques is the multi-sphere approach, which involves rigidly connecting a group of spheres to create irregular particles \cite{abou2004three,kodam2009force,cabiscol2018calibration}. In the multisphere approach, each individual subsphere serves the purpose of contact detection, enabling an accurate interaction between particles. However, when it comes to calculating inertial characteristics and integrating particle motion, the entire collection of spheres is considered as a single particle. This simplification reduces computational complexity and facilitates efficient simulations. The fidelity of the multisphere particle's morphology, or how closely it approximates the shape of the target particle, is determined by the number and size of the spheres used in its construction. A higher number of spheres or smaller sphere sizes yield greater morphological fidelity, providing a better representation of the target shape. However, this also increases the computational cost associated with interactions between neighboring particles, as more contact points need to be considered. Therefore, the choice of the number and size of spheres in a multisphere particle involves a trade-off between morphological accuracy and computational efficiency. Researchers must carefully consider these factors based on the specific requirements of their simulation to strike an appropriate balance. Overall, the multisphere approach offers a flexible and widely used method for modeling non-spherical particles within DEM simulations, providing a compromise between accuracy and computational efficiency.
\begin{figure}[htb!]
    \centering
    \begin{subfigure}[t]{0.49\textwidth}
    \centering 
\includegraphics[width=\textwidth]{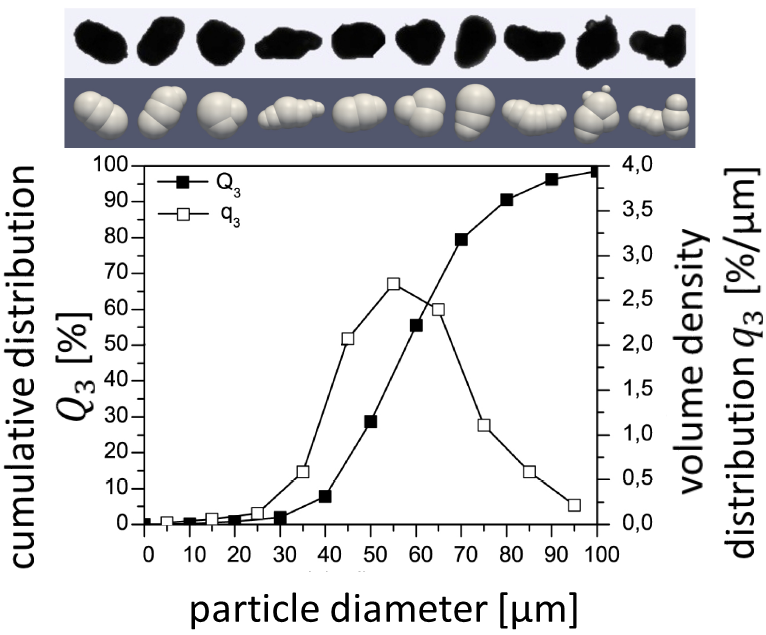}
\caption{}
     \end{subfigure}
     \hfill
   \begin{subfigure}[t]{0.49\textwidth}
    \centering   
\includegraphics[width=\textwidth]{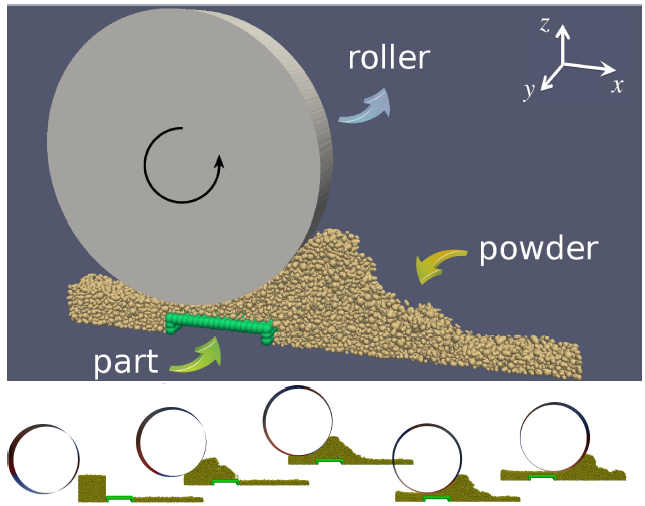}
\caption{}
   \end{subfigure}
\caption{(a) (Top) Light microscope images of commercially available PA12 powder particles (first row) and corresponding
particle models using the multisphere method (second row) for implementation in the DEM. (Bottom) Cumulative distribution and volume density distribution as a function of the particle diameter. (b) Snapshot of the simulation indicating the main elements of the powder application process.}
\label{fig:powderspreading}
\end{figure}

\autoref{fig:powderspreading}(a) shows some examples in which target shapes were obtained from SEM images of commercially available PA12 powder as described in \cite{parteli2016particle}. The translational and rotational motion of the multisphere particles is driven by the forces acting on the constituent spheres, which include gravity, inter-particle forces, and the forces due to particle and wall interactions. The small size particles ($1-100\,\upmu{}$m) involved in the process require attractive interaction forces to be modeled and incorporated into the DEM. We developed a particle-based model which takes into account contact forces, as well as inter-particle attractive interactions. The contact forces are modeled using the Cundall and Strack model \cite{cundall1979discrete} considering elastic and dissipative forces in the normal collision direction and in the tangential direction. This model was extended by 
Parteli et al. \cite{parteli2014attractive} to incorporate attractive particle interaction forces. This improved DEM model takes into account bonded adhesion contacts as well as non-bonded van der Waals forces. The simulation predictions of the solid fraction of powders covering a wide range of size distributions were quantitatively in agreement with the experimental results \cite{parteli2014attractive}.

\autoref{fig:powderspreading}(b) shows the simulation snapshot indicating the main components of the powder spreading setup implemented for numerical simulation using the DEM model. The roller moves in the positive $x$ direction with a translational velocity because it rotates in a counterclockwise direction with a constant rotational velocity.

\subsection{Influence of particle size}
\begin{figure}[htb!]
    \centering
\includegraphics[width=0.5\textwidth]{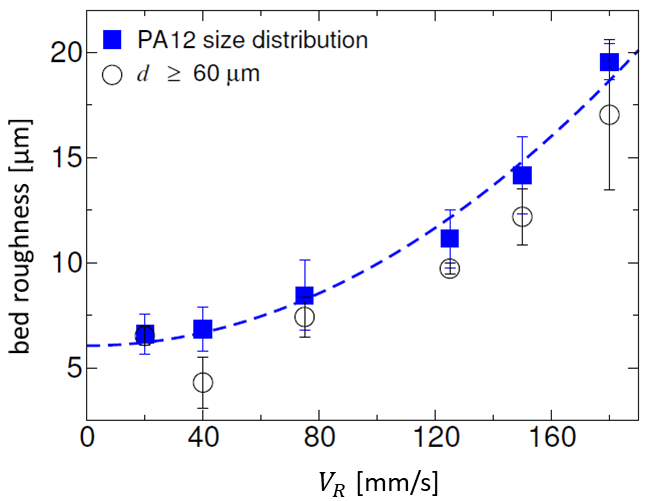}
\caption{Roughness of the deposited powder layer as a function of the translational velocity of the roller $V_R$ \cite{parteli2016particle,parteli2017particle}.}
\label{fig:roughness} 
\end{figure}

The packing behavior of the deposited powder layer is strongly dependent on the transport dynamics of the powder over the substrate. In ideal powder spreading conditions, the goal is to enhance the production speed while maintaining flat, densely packed powder layers. To investigate this behavior, we analyzed the roughness of the powder bed as a function of the roller velocity $V_R$ as shown in \autoref{fig:roughness}. We observe a clear increase in the bed roughness with the increase in the roller's translational velocity as represented by the blue squares in \autoref{fig:roughness}.  

To further analyze how the particle size distribution affects the roughness of the bed, we modified the size distribution shown in \autoref{fig:powderspreading}(b) by removing all particles of diameter smaller than $60\,\upmu$m. The surface roughness resulting from particles of size $d \geq 60\,\upmu$m is shown as a function of the roller's translational velocity, represented by the circles in \autoref{fig:roughness}. Overall, we achieve a smaller roughness with the larger particle size distribution, and the roughness increases in the same fashion with the roller's velocity as it does for the original size distribution.

Note that the attractive inter-particle forces (adhesion and non-bonded van der Waals forces) lead to the formation of large agglomerates in the powders associated with the blue squares in \autoref{fig:roughness}. For very small particles $< 30\,\upmu$m, the gravitational forces are dominated by the cohesive forces, and thus, the particles are transported as irregular clumps and deposited on the surface. This results in a higher roughness than the original size distribution of the powder. Thus, a way to decrease the surface roughness of the deposited powder layers in additive manufacturing is to modify the particle size distribution of the powder by filtering out the smaller particles from the system.

\subsection{Influence of particle shape}

\begin{figure}[htb!]
    \centering
\includegraphics[width=0.8\textwidth]{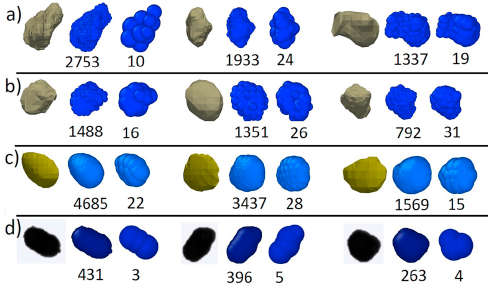}
\caption{Multisphere reconstruction of particles used in additive manufacturing. From left to right: original template (SEM image), all spheres before optimization, and optimized multisphere representation. (a) S1 sample (PA11), (b) S2 sample (PA11 powder rounded through precipitation), (c) S3 sample (PA12), and (d) S4 sample (PA12 reconstructed from SEM images) \cite{nasato2020influence,NasatoHeinlHausottePoeschel:2017}.}
\label{fig:multispherereconstruction} 
\end{figure}

The role of real particle shape on the powder layer quality has been explored in detail by Nasato et al. \cite{nasato2020influence}. This is done by multisphere reconstruction of 3D images of the powder particles obtained using the CT-REX tomograph as shown in \autoref{fig:multispherereconstruction}. Samples from commercial polyamide 11 (PA11), modified PA11, polyamide PA12, and reconstructed images from PA12, named S1, S2, S3, and S4, respectively, are considered for this study. 

\begin{figure}[htb!]
    \centering
    \begin{subfigure}[t]{0.49\textwidth}
    \centering 
\includegraphics[width=\textwidth]{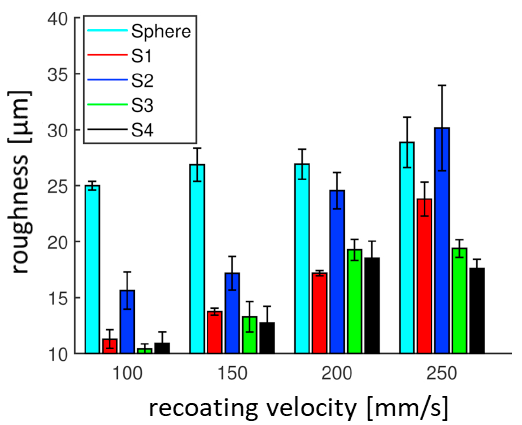}
\caption{}
     \end{subfigure}
     \hfill
   \begin{subfigure}[t]{0.49\textwidth}
    \centering   
\includegraphics[width=\textwidth]{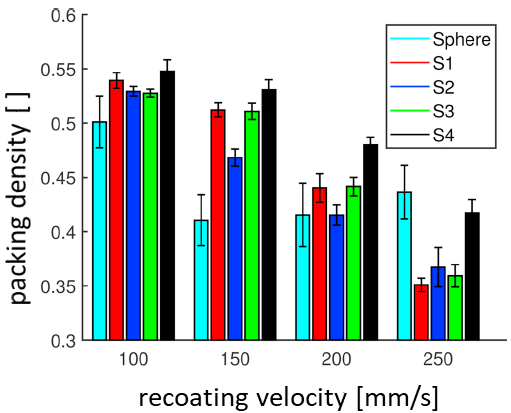}
\caption{}
   \end{subfigure}
\caption{(a) Roughness and (b) packing density of the deposited powder layer as a function of the translational velocity of the roller $V_R$.}
\label{fig:PAroughness} 
\end{figure}

The surface roughness is measured for all four samples for different recoating velocities of the powder spreading tool. Interesting to note from \autoref{fig:PAroughness}(a) that for every recoating velocity, the surface roughness is the highest for either the spheres or the sample S2 (rounded PA11). Samples S1, S3, and S4 have similar results of roughness within the standard deviation. Note that samples S3 and S4 have a smaller roughness than sample S1, particularly for the higher recoating velocity of 250\,mm/s. Samples S3 and S4 have the lowest value of aspect ratio and elongation ratio that is measured. Thus, it is speculated that the smaller surface roughness of S3 and S4 samples is related to the alignment of the particles during the powder spreading process, leading to a smoother surface. An analysis of the packing density results for the different powder samples is shown in \autoref{fig:PAroughness}(b). For up to recoating velocity of 200\,mm/s, samples S1 to S4 achieve higher packing densities than the sphere sample. The sample using spheres reveals a higher packing density only for the high recoating velocity of $250$\,mm/s. This is explained by the better flowability of the spheres at high velocities compared to the non-spherical particles. Thus, a competitive mechanism influences the quality of the powder layer in the recoating process. Spherical particles generally have better flowability and are thus preferred for additive manufacturing applications. However, this holds true only for high recoating velocities. For lower recoating velocities, the elongated shape of the particles plays a dominant role in maintaining higher packing densities.

\section{Attractive interactions between particles}
It is crucial to consider not only repulsive contact forces but also attractive forces arising from van der Waals interactions when modeling the dynamics of powder particles accurately. Van der Waals forces are a type of intermolecular force that results from fluctuations in electron distributions, creating temporary dipoles and inducing attractive forces between particles. In this section, our focus is on investigating the effects of quantum mechanically based van der Waals forces and tensile forces arising from surface adhesion on the macroscopic behavior of powder particles on the packing structure of powder. While van der Waals forces are generally associated with intermolecular interactions at the atomic or molecular level, their influence can extend to macroscopic particles of small size range \cite{parteli2014attractive}.

\subsection{Significance of attractive forces}
\begin{figure}[htb!]
    \centering
    \begin{subfigure}[t]{0.44\textwidth}
    \centering 
\includegraphics[width=\textwidth]{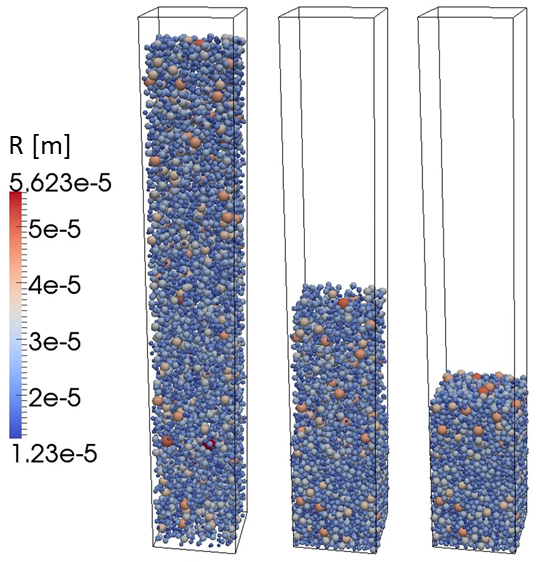}
\caption{}
     \end{subfigure}
     \hfill
   \begin{subfigure}[t]{0.55\textwidth}
    \centering   
\includegraphics[width=\textwidth]{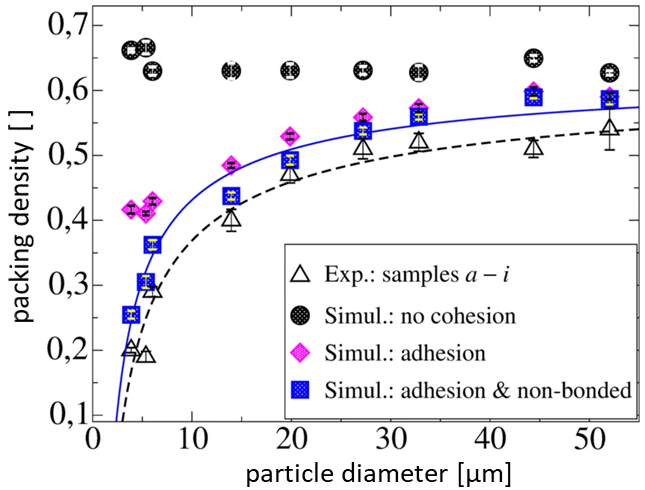}
\caption{}
   \end{subfigure}
\caption{(a) Numerical simulation of the powder packing of size distribution corresponding to sample $i$ with radius varying from $10-55\,\upmu$m and $R$ representing the radius of the particles and (b) packing density as a function of the average particle diameter.}
\label{fig:packingfraction} 
\end{figure}

In cooperation with experimental validation, it must always be verified that the models used for DEM simulations lead to results that are consistent with the physical reality \cite{schmidt2020packings}. Therefore, the numerical tool, which is extended to include a more detailed description of particle geometry and the electrostatic inter-particle interactions, is first tested for simulations of a granular system in a simple geometry by Parteli et al. \cite{parteli2014attractive}. They verified the bulk packing density of fine glass powders in the diameter range $(4-52)\,\upmu$m by comparing the results with experiments and DEM simulations. The numerical setup is demonstrated in \autoref{fig:packingfraction}(a), where silica glass beads are deposited in a rectangular box. We adopt periodic boundary conditions in the $x$ and $y$ directions and a frictional wall at $z=0$. The height of the box is considered large enough to produce packings of depth larger than $30\langle d\rangle$. To validate our models, we obtained qualitative agreement between the experimental and numerical results, taking into account attractive forces for particle interaction, which include adhesion force and non-bonded van der Waals forces. 

We performed different sets of simulations assuming different particle interaction force models, namely (i) viscoelastic interaction (no cohesion in \autoref{fig:packingfraction}(b)), (ii) adhesive and viscoelastic interaction (adhesion in \autoref{fig:packingfraction}(b)), and (iii) full model, including viscoelastic, adhesive, and van der Waals interaction (adhesion and non-bonded in \autoref{fig:packingfraction}(b)). The results are shown in \autoref{fig:packingfraction}(b). The packing density obtained from the experimental results was compared to the given sets of simulations. For the pure viscoelastic interaction model, the obtained packing density is almost independent of the mean particle size, $\langle d\rangle$, with a slight tendency of increasing packing density for smaller particles. This may be explained by the geometric effects of smaller particles filling the pore space. The simulation results with the viscoelastic interaction models agree with experimental results for large particle size, $\langle d\rangle \gtrsim 40\,\upmu$m but disagree for smaller particle size. Simulations incorporating JKR-type adhesive forces reveal a decay in $\phi$ with decreasing $\langle d\rangle$. A good agreement is obtained with the experiment for $\langle d\rangle \gtrsim 20\,\upmu$m, but the data diverge for smaller particles. Simulations with the full model, including viscoelastic, JKR-adhesive, and van der Waals interaction, allowed us to reproduce the experimentally measured packing fraction for the entire particle size interval $4\,\upmu\mathrm{m} \lesssim \langle d\rangle \lesssim 52\,\upmu$m. Thus, we conclude from this that, for a predictive simulation of fine powder behavior, both adhesive and van der Waals forces are essential and should, thus, be considered in DEM simulations. Neglecting any of these contributions in simulations of fine powders may lead to unreliable results.

\subsection{Effect of inter-particle cohesion on packing structure}
Inter-particle cohesion plays a key role on the structure and density of the deposited powder layer, and thus, in the overall quality of the end product \cite{he2020linking,roy2023effect}. Recently, we explored the structural anisotropy, packing density, and surface profile roughness of the deposited powder layer in spreading processes with varying cohesion using DEM simulations \cite{roy2022local}. The local density and surface roughness of the produced layers were also quantified to identify the influence of cohesion on the formation of surface texture. Remarkably, based on the statistics of the local anisotropy and the surface roughness, we observe a synchronized transformation of the structural anisotropy and the surface roughness profile with increasing cohesion. In particular, the structure changes from homogeneous to heterogeneous, the layer surface roughness increases, and its peaks change shape, around $Bo\approx10$, where $Bo$ is a measure of inter-particle cohesion strength in dimensionless form. In the following subsections, we highlight our main observations based on the aforementioned characteristics of powder layer packings.

\begin{figure}[htbp]
  \begin{center}
    \includegraphics[width=\columnwidth]{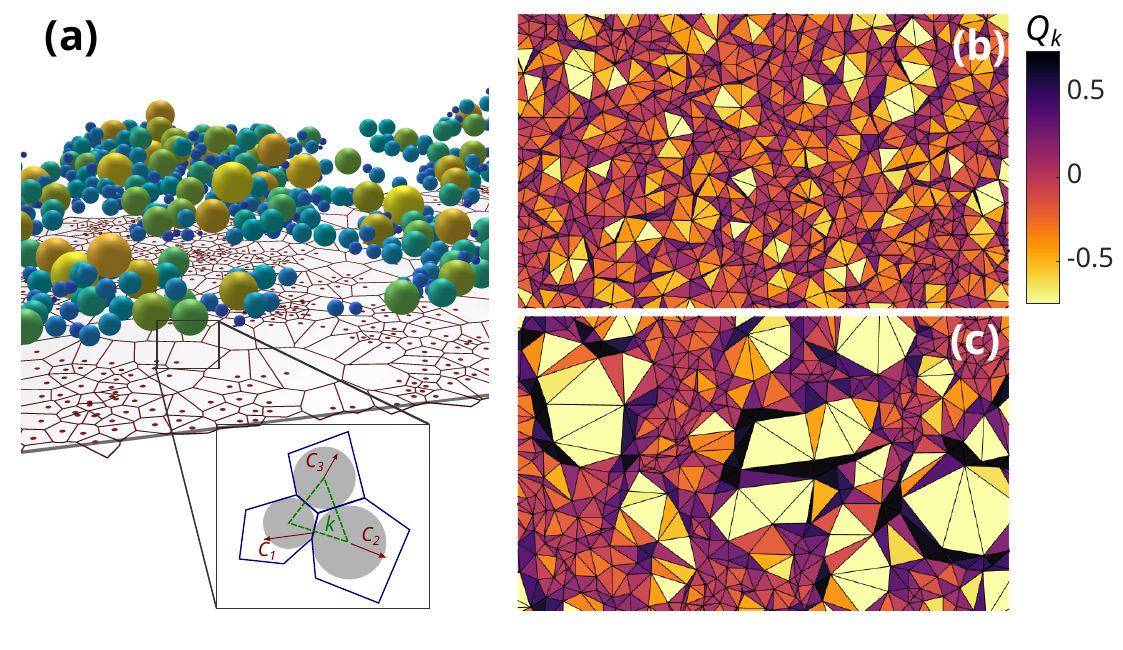}
  \end{center}
\caption{Characterizing the structural anisotropy in the deposited layer. (a) Granular packing of the powder layer for $Bo\approx30$. The points represent the projections of the center points of the particles in the plane. The corresponding 2D Voronoi tessellation is also shown. Also shown in (a) is a schematic representation of the particle packing with the superimposed Voronoi tessellation (blue) and Delaunay triangles (green). The vectors $C_p$ (red) point from the center of the particles to the centroids of the Voronoi cells. The calculated $Q_k$ for (b) $Bo=0$ and (c) $Bo\approx30$, where the Delaunay triangles are colored by the corresponding values $Q_k$. }
\label{fig:voronoi}
 \end{figure}

\autoref{fig:voronoi}(a) illustrates an exemplary deposited layer of highly cohesive particles, characterized by the Bond number $(Bo\approx30)$. On examination, we observe a heterogeneous structure comprising both densely packed and loosely packed regions. This variation in packing density highlights the complexity and non-uniformity of particle arrangements within the deposited layer. The presence of regions with varying packing densities within the deposited layer of highly cohesive particles renders the overall packing density inadequate as a sole characterization metric. In order to address the challenge of characterizing the structural inhomogeneity in the deposited layer, we propose a novel method that utilizes a threshold-free measure based on the local structural anisotropy. The method of characterizing the structural inhomogeneity using a threshold-free measure based on the local structural anisotropy, measured by $Q_k$, and calculated from Voronoi tessellation, is explained in \cite{roy2022local}. 
As stated by Rieser et al. (2016) \cite{rieser2016divergence}, in regions of dense and homogeneous packing, the values of $Q_k$ tend to exhibit random fluctuations, resulting in a Gaussian-shaped distribution. However, in regions characterized by inhomogeneity and high anisotropy, the distribution of $Q_k$ deviates from the Gaussian shape. In such cases, both highly positive and highly negative values of $Q_k$ appear at the tails of the distribution.

\begin{figure}[t]
    \begin{center}
        {\includegraphics[width=\columnwidth]{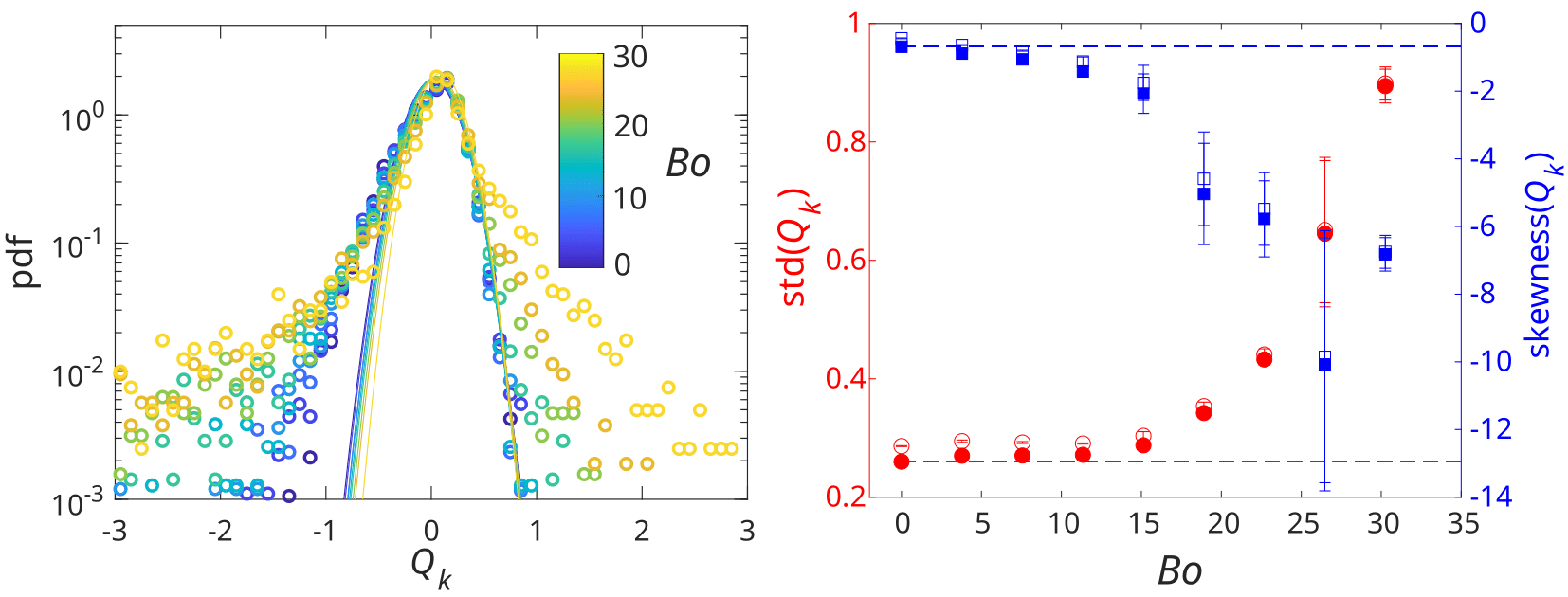}}\\
    (a)\hspace*{5cm}(b)
    \end{center}
    \caption{(a) Probability density of normalized divergence of center-to-centroid vectors for the quasi-2D packing of powder deposited for different $Bo$. $Q_k  > 0$ regions are more densely packed than their surroundings; hence, we call these regions overpacked. $Q_k < 0$ regions are more loosely packed than their surroundings and are therefore labeled underpacked. (b) Standard deviations (red circles) and skewness (blue squares) vs. $Bo$. We also compared the standard deviations and skewness of the distributions with the divergence calculated from the condition without 2D projections (hollow points) and the condition with 2D projections (solid points), as explained in \cite{roy2023structural}. Error bars reflect standard errors in three simulation repetitions. The dashed lines indicate the standard deviation and the skewness of the distribution $Q_k$ for $Bo=0$. }
    \label{fig:pdf}
\end{figure}

Figure \ref{fig:pdf}(a) 
shows the distribution of $Q_k$ for different Bond numbers ($Bo$), providing visual evidence of this behavior. The non-Gaussian shape observed in the distribution indicates the presence of inhomogeneous regions with pronounced anisotropy, where the particle arrangements deviate significantly from random fluctuations. This analysis enables the characterization and differentiation of regions with varying degrees of structural complexity, providing valuable insights into the packing behavior as influenced by the Bond number. Referring to \autoref{fig:pdf}(b), we observe interesting transition behavior of the standard deviation and skewness of the $Q_k$ distribution for different Bond numbers ($Bo$). Initially, up to $Bo\approx10$, the standard deviation remains relatively constant, indicating a consistent level of structural anisotropy. Similarly, the skewness also exhibits minimal variation within this range. However, beyond $Bo\approx10$, both the standard deviation and skewness start to exhibit a monotonic increase and decrease, respectively. This suggests a significant shift in structural anisotropy for cohesive materials, with $Bo\approx10$ serving as a transition point. For higher cohesion values $(Bo > 10)$, particles have the ability to maintain larger voids during the spreading process, resulting in a more pronounced anisotropic structure. The observed fluctuations in the trends for higher $Bo$ values, even within repeated simulations, can be attributed to the inherent high anisotropy of the structures. These fluctuations highlight the complexity and variability of particle arrangements in cohesive systems with enhanced anisotropy. By analyzing the standard deviation and skewness of the $Q_k$ distribution, we gain valuable insights into the transition in structural anisotropy for cohesive materials and the impact of different Bond numbers on the packing behavior.

\section{Thermal and mechanical properties}
This section aims to expand our understanding of the various aspects involved in powder application and gain a more comprehensive insight into the process dynamics by combining the thermal and mechanical effects of the process and material parameters on the powder application process. To incorporate heat transfer capabilities into our Discrete Element Method (DEM) model for simulating non-spherical particles, we have extended the existing framework by merging a multisphere algorithm with a thermal discrete particle model \cite{ostanin2023rigid}. The heat transfer for multi-sphere particles includes conduction, convection, and radiation. Conduction occurs when a clump overlaps with another clump or a boundary, while convection and radiation occur for the exposed portion of a clump's surface with the background. The total thermal transfer rate is expressed as the sum of the heat flux from conduction, convection, and radiation. To demonstrate the effectiveness of our extended DEM model with heat transfer capabilities, we conducted simulations of a powder spreading process using Polyamide 12 (PA12) powder particles of irregular shapes. 

\subsection{Multi-sphere particle generation}
In this study, multi-sphere particles were generated to approximate the shape of Polyamide 12 (PA12) powder particles. To achieve this, an image-informed particle generation procedure was implemented, utilizing scanning electron microscopy (SEM) images of PA12 particles as a reference \cite{nasato2021influence}. The procedure involved extruding the SEM images into three-dimensional (3D) representations, which served as the target geometry for generating realistic multi-sphere particles. The generation of multi-sphere particles was performed using a recently developed method based on the Euclidean transform of 3D images. The open-source software \texttt{CLUMP} \cite{angelidakis2021clump} was employed, which offers various techniques for the multisphere generation. For adequately approximating the target particle shapes, a total of 10 sub-spheres were used per particle. This choice of the sub-sphere count was determined to strike a balance between accurately representing the target particle morphology and maintaining computational efficiency. By employing the image-informed particle generation procedure and the multi-sphere approach with 10 sub-spheres per particle, the study aimed to create realistic representations of PA12 powder particles for subsequent simulations and analysis. The different irregular particle shapes that are included in our studies are from light microscope images of commercially available PA12 as illustrated in \autoref{fig:powderspreading}(a).

\subsection{Simulation of powder spreading of multisphere particles with thermal properties}

\begin{figure}[htb!]
    \centering
    \includegraphics[width=1.0\textwidth]{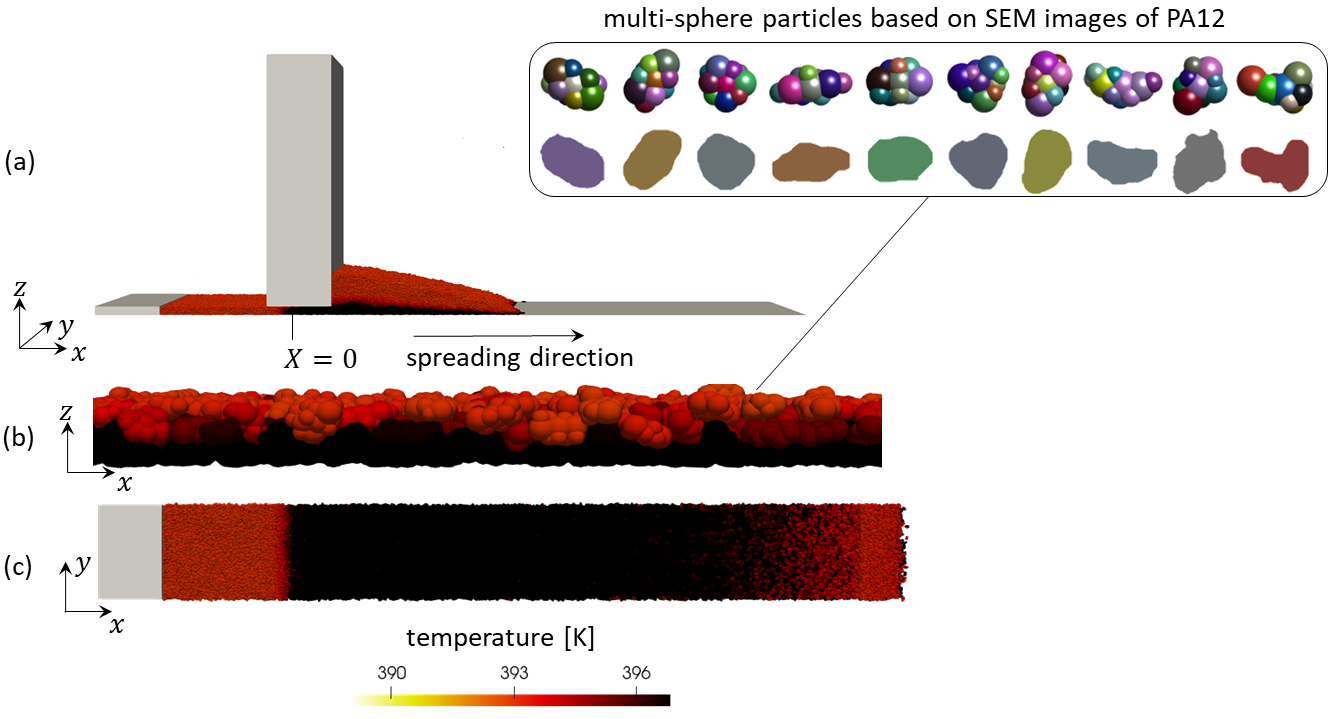}
    \captionsetup{justification=justified}
    \caption{(a) Powder spreading setup with a blade speed $v = 10$\,mm/s, showing (b) front view after $0.08$\,s and (c) top view after $1.0$\,s of the spread powder layer.}
    \label{fig:powderSpreading}
\end{figure}

A small part of the powder bed (width $0.75$\,mm) is simulated using periodic boundary conditions in the $y$-direction. The contact model used for this simulation is Hertz-Mindlin model combined with viscous damping \cite{muller2011collision}. In this numerical setup, the substrate is assumed to be smooth. We insert multispheres of 10 different irregular shapes generated from images of commercially available PA12 particles as shown in \autoref{fig:powderSpreading}. The multispheres are inserted in front of the spreader tool until the total bulk particle volume equals $0.7$\,mm$^3$, which is sufficient material to create a powder layer of $5$\,mm length, $0.75$\,mm width and $100\,\upmu$m height. We allow the particles to settle and relax under the effect of gravity before we start the spreading process. These simulations begin with particle temperatures set at $T_\mathrm{part} = 393$\,K, and the powder is spread over a pre-heated plate held at $T_\mathrm{plate} = 430$\,K, as previously documented in \cite{li2021experimental,laumer2015influence,peyre2015experimental}. After the particles have settled down and the system is relaxed, the spreading process starts by moving the tool at a constant speed $v$\,mm/s. Particles reaching the end of the powder bed (at $x = 5$\,mm) are not used for the analysis. Finally, the simulation is stopped when the system is static, i.e., the kinetic energy of the system is sufficiently low. Thus, the maximum simulation time is chosen as  $t_\mathrm{max} = 1$ s, taking into account the time for initial settling and relaxation of particles, the spreading time, and the time for the system to reach a static state. A detailed description of the setup can be found in \cite{roy2023thermal}.

\subsection{Packing density analysis with thermal influence}
The elastic modulus is dependent on the temperature as \cite{luding2005discrete}

\begin{align}
&E^*(T_{ij}) = \frac{E_0^*}{\alpha}\bigg[1+(\alpha-1)\tanh\bigg(\frac{T_m-T_{ij}}{\Delta T}\bigg)\bigg]
\label{eq:functionModulus}
\end{align}
where $T_m = 451$ K is the melting temperature, $\Delta T = 20$ K is the temperature interval over which phase change occurs, and $\alpha$ is the parameter defining the dependency of the elastic modulus on the temperature. The $E^*(T_{ij})$ model proposed by Luding et al. \cite{luding2005discrete} conforms to $\alpha = 2$, which needs further experimental validation. Note that according to \autoref{eq:functionModulus}, an extreme softening of the material of up to $80\%$ occurs at a large value of $\alpha$ and at an extreme temperature of the particles, close to their melting point. However, in our powder spreading setup, such a large temperature is not achieved by particles without sintering.

\begin{figure}[htb!]
    \centering
    \begin{subfigure}[t]{0.49\textwidth}
    \centering 
    \includegraphics[width=\textwidth]{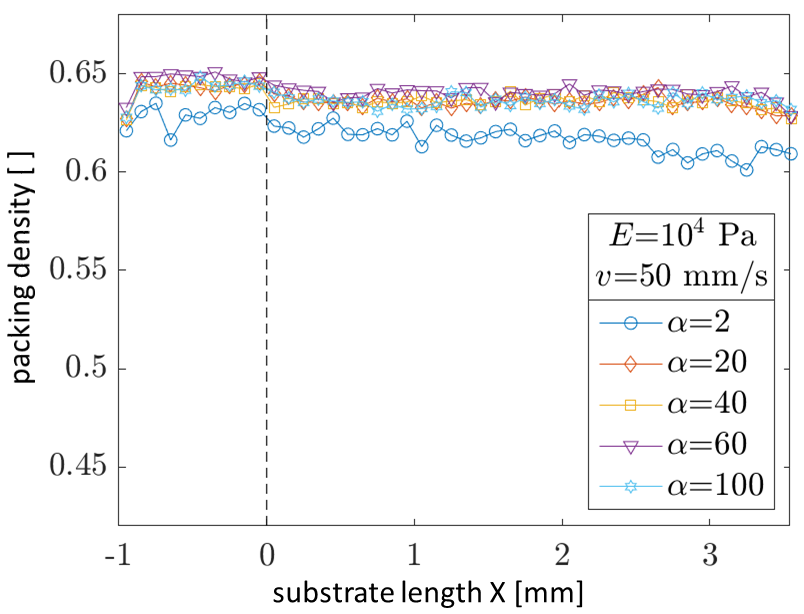}
\caption{}
     \end{subfigure}
     \hfill
   \begin{subfigure}[t]{0.49\textwidth}
    \centering   
    \includegraphics[width=\textwidth]{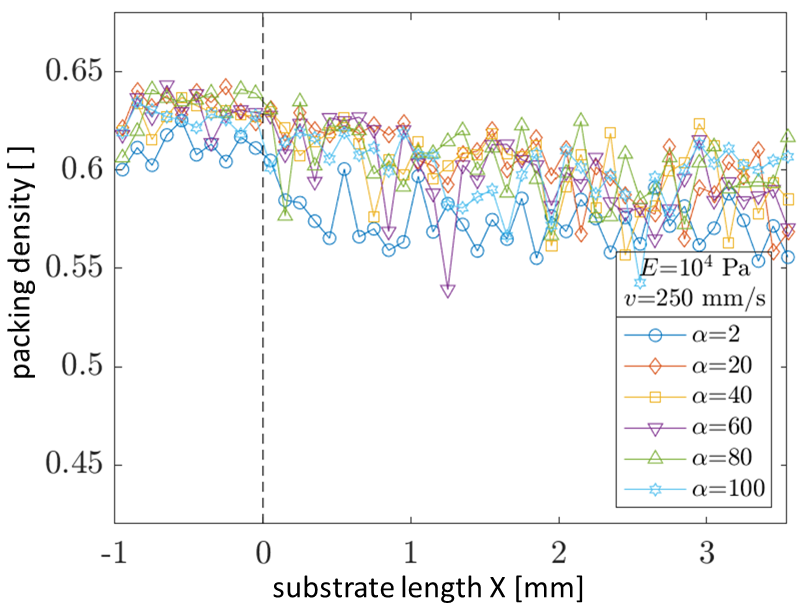}
\caption{}
   \end{subfigure}
\caption{Packing density as a function of substrate length $X$ for different coefficient $\alpha$ for (a) $v = 50$\,mm/s and (b) $v = 250$\,mm/s for initial elastic modulus $E_0^* = 10^4$ Pa.}
\label{fig:modulustemperaturedependence}
\end{figure}

We characterized the packing density of the powder layer subjected to varying blade velocity and the dependence of the particle stiffness on temperature. In \autoref{fig:modulustemperaturedependence}, we observe that the packing density is significantly influenced by the blade velocity. The packing density corresponding to a lower blade speed of 50\,mm/s in \autoref{fig:modulustemperaturedependence}(a) is lower than that in \autoref{fig:modulustemperaturedependence}(b) for blade speed of $250$\,mm/s. Further, the packing density is more sensitive to the values of $\alpha$ at a lower blade speed of $v = 50$\,mm/s. Note that the packing density is more sensitive to a smaller value of $\alpha$, i.e., there is a significant difference in packing fraction between $\alpha = 2$ and $20$ than between the values for $\alpha > 20$. A lower blade speed allows more time for inter-particle heat exchange, and thus the value $\alpha$ plays a significant role in softening the material. This leads to a change in the packing density of the powder layer.

\begin{figure}[htb!]
    \centering
    \begin{subfigure}[t]{0.49\textwidth}
    \centering 
    \includegraphics[width=\textwidth]{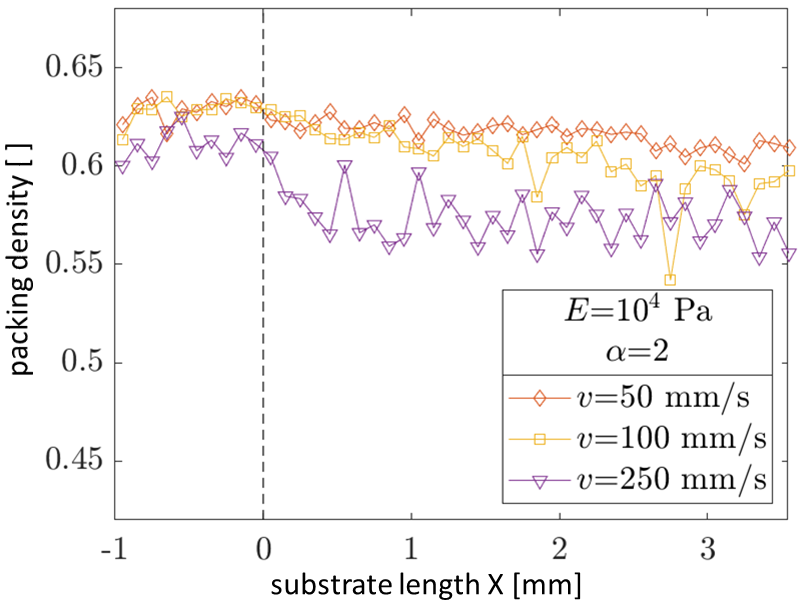}
\caption{}
     \end{subfigure}
     \hfill
   \begin{subfigure}[t]{0.49\textwidth}
    \centering   
    \includegraphics[width=\textwidth]{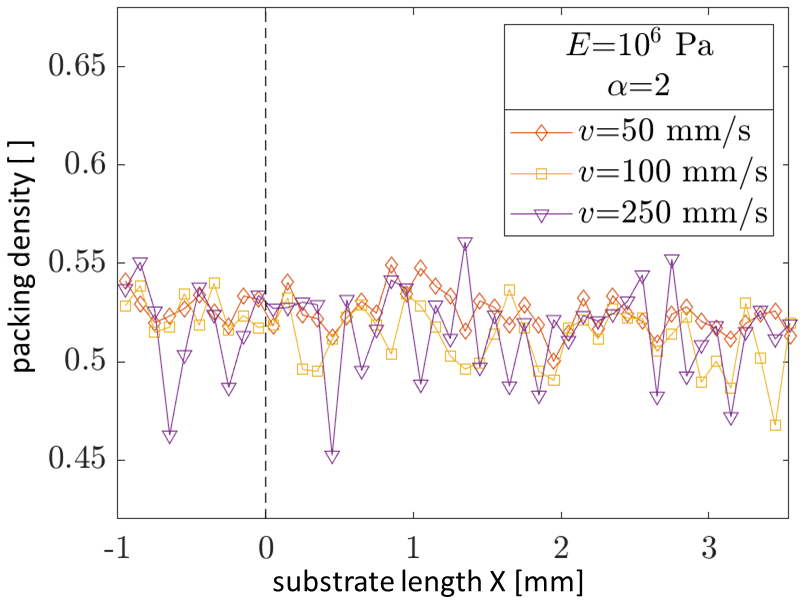}
\caption{}
   \end{subfigure}
\caption{Packing fraction for different blade speed $v$ for different initial elastic modulus (a) $E_0^* = 10^4$ Pa and (b) $E_0^* = 10^6$ Pa for $\alpha = 2$.}
\label{fig:bladespeed}
\end{figure}

In \autoref{fig:bladespeed}(a) 
 and (b), we compare the packing density for different blade speeds. The real modulus of PA12 powder is on the order of $10^8$ Pa. However, for the sake of computational efficiency, we use a lower elastic modulus in our simulations. To further test the effect of the elastic modulus on the packing density of the powder layer, we simulate the powder spreading process using two different values of the initial elastic modulus, $E_0^*$: $10^4$ Pa and $10^6$ Pa, respectively, while keeping $\alpha = 2$ as constant. We observe that the packing density is less sensitive to blade velocity for material with higher elastic modulus. This is because stiff particles allow less interparticle compression, and thus the packing density is less sensitive to the blade speed. For soft particles, the packing density is influenced by the flow dynamics of the material, which depends on the blade speed.

\section{Vibrating recoating mechanism}
\begin{figure}[htb!]
    \centering
    \begin{subfigure}[t]{0.49\textwidth}
    \centering 
    \includegraphics[width=\textwidth]{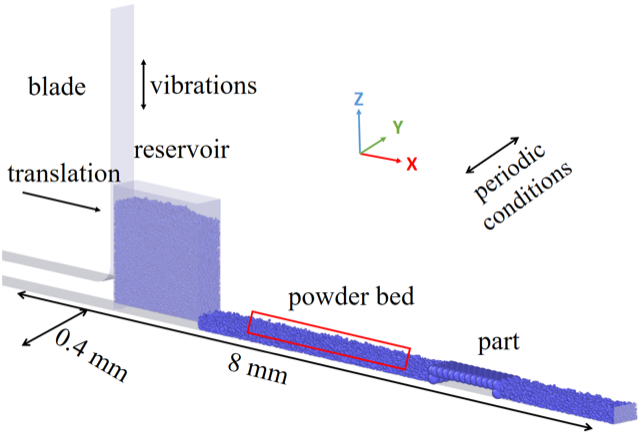}
\caption{}
     \end{subfigure}
     \hfill
   \begin{subfigure}[t]{0.49\textwidth}
    \centering   
    \includegraphics[width=\textwidth]{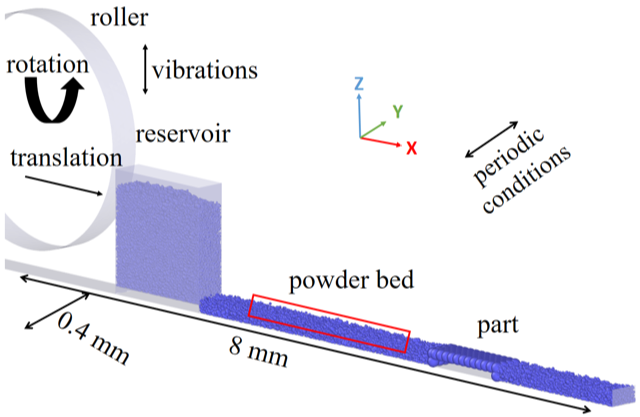}
\caption{}
   \end{subfigure}
\caption{Powder spreading simulation setup using (a) a blade and (b) a roller recoating tool. Red boxes indicate the powder bed region where porosity is evaluated after the layer is recoated. Porosity is additionally evaluated on the top of the part (source \cite{nasato2021influence}).}
\label{fig:bladeroller}
\end{figure}

Improving the layer quality can be achieved via modification of the powder application process \cite{marchais20213d,haeri2017discrete}. Nasato et al. (2021) \cite{nasato2021influence,NasatoPoeschelParteli:2016} explored the applicability of vibrating recoating mechanisms as a technique that can be tailored according to material characteristics, such as particle size, shape, and cohesion, to produce layers of improved characteristics. They performed DEM simulations using cohesive, multisphere particles of realistic shapes of PA12 particles to explore powder spreading systems using two spreading tools, namely a blade and a roller. \autoref{fig:bladeroller} shows the simulation system.

Vibration applied on the PA12 material via the recoating tool was used as a means of compacting the powder particles during their application to form the layered bed. The influence of the vibration characteristics on the layer density and surface roughness was investigated. The vibration frequency was varied to values of 100, 250, 500, and 1000\,Hz, while the amplitude was varied to 2, 5, 10, and 20 \textmu{}m, for recoating velocities of 100, 150, 200, and 250\,mm/s.

\begin{figure}[htb!]
    \centering
    {\includegraphics[width=0.99\textwidth]{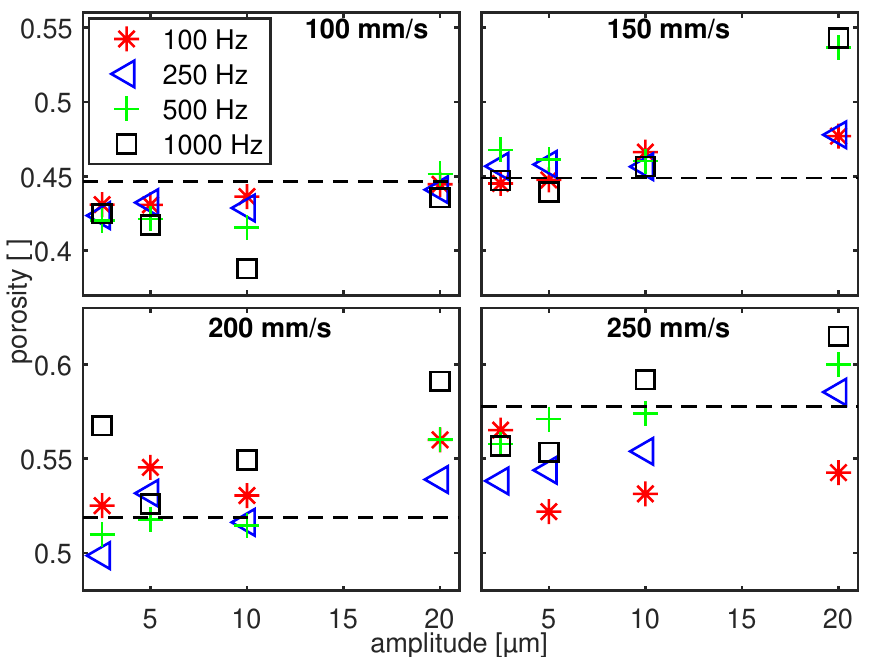}}
\caption{Porosity of granular layer obtained for different amplitudes (x-axis) and frequencies. The powder spreading tool is a blade, operating with a translational velocity from 100\,mm/s (top left) to 250\,mm/s (bottom right). The dashed line indicates the reference case where no vibration was applied. The porosity was measured on the powder recoated on the top of the granular bed, see also \cite{nasato2021influence}).}
\label{fig:vibratiionporosity}
\end{figure}
\autoref{fig:vibratiionporosity} demonstrates that for recoating velocity 100\,mm/s, considering large frequency (i.e., 1000\,Hz) combined with intermediate amplitude (10 \textmu{}m) leads to a major improvement of the powder bed density, reducing the porosity by 13.2\%. However, for larger amplitudes, no significant change in the porosity is observed. This indicates an optimum between 10 and 20 \textmu{}m for the amplitude. In all cases, increased vibration energy seems to disturb the powder layer, and leads to a loosening of the granular bed, as seen e.g., for 1000\,Hz and 20\,\textmu{}m. This effect can be observed also for recoating velocity 150\,mm/s, where the porosity of the granular bed increases for large frequencies and amplitudes by 20.9\% for 1000\,Hz and 20 \textmu{}m.

\begin{figure}[htb!]
{\includegraphics[width=0.8\textwidth]{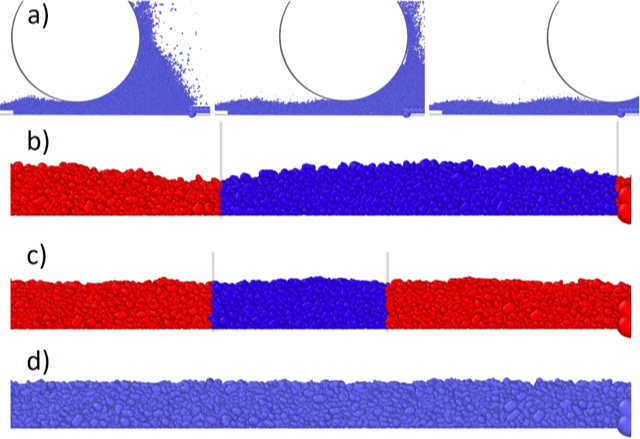}}
\caption{Formation of bumps on the surface of the powder bed as a result of the vibrating mechanism. (a) Snapshots taken at different stages of the recoating process show the formation of the bumps for (b) recoating velocity 250\,mm/s, frequency 100\,Hz and amplitude 20 \textmu{}m (c) recoating velocity 100\,mm/s, frequency 100\,Hz and amplitude 20\,\textmu{}m. $D_b$ is 2.5\,mm and 1.0\,mm for cases (b) and (c), respectively. The particles marked in dark blue indicate a region between the periodic bump, and correspond to $D_b$. (d) No bump could be identified for recoating velocity 100\,mm/s, frequency 250\,Hz and amplitude 20\,\textmu{}m. (source \cite{nasato2021influence}).}
\label{fig:layerbumps}
\end{figure}

The influence of vibrations on the recoating mechanism
using a roller was investigated by \cite{nasato2021influence}. They found that vibrations increased the porosity of the recoated layer in all cases for recoating velocities above 150\,mm/s, which they attributed to being an artifact of the roller geometry, largely influencing the powder bed. The roller introduces an artificial bumpiness on the layer surface, as shown in \autoref{fig:layerbumps}, which is, however, minimized for small recoating velocities, such as 100\,mm/s.

\section{Conclusions}
We developed a numerical tool specifically designed for particle-based DEM simulations in the context of powder application within additive manufacturing devices. Our tool takes into account the intricate geometric shapes of the particles involved. To assess the capabilities of our model, we utilized it to investigate the transport of powder particles using a roller as coating system. Through our simulations, we were able to discern that the process velocity has a significant influence on the packing characteristics of the applied powder. The findings of our study highlight the importance of process speed in determining how the powder particles pack together during the coating process. This understanding can be valuable in optimizing the additive manufacturing process, allowing for improved control over the final properties and quality of the manufactured objects. By incorporating complex particle geometric shapes into our simulations, our numerical tool provides a more realistic representation of the powder application process. This enables us to gain insights into the underlying mechanisms governing particle transport and packing, aiding in the development of more efficient and effective additive manufacturing techniques.

Subsequently, we expand upon our existing Discrete Element Method (DEM) model by introducing inter-particle attractive forces. We emphasized the importance of considering two essential contributions to the particle interaction model: adhesion and van der Waals forces. Adhesion is incorporated into our model using the Johnson-Kendall-Roberts (JKR) theory, while van der Waals interactions are also taken into account. We assert that neglecting either of these contributions leads to significant deviations in the simulation results, particularly for fine powders. Such deviations render the DEM method unreliable in accurately capturing the packing behavior.

Furthermore, our investigation considered the influence of powder cohesion on the structural anisotropy of the deposited layers. Structural anisotropy refers to the variation in the spatial distribution of particles within the layers. It provides insights into the overall organization and packing behavior of the powder particles during the spreading process. Understanding the effect of powder cohesion on structural anisotropy is important for optimizing processes where particle arrangement plays a significant role, such as in additive manufacturing, powder compaction, or coating applications. 

Finally, the focus of the research was to further explore the powder application process through process simulations. In particular, the thermal and mechanical effects of the process and material parameters on the powder application process are the primary research objectives. The integration of the multisphere algorithm with the thermal discrete particle model in the DEM framework done during this phase of research represents a significant advancement in simulating non-spherical particles with heat transfer capabilities. This extension opens up new possibilities for studying a wide range of systems, such as granular materials, powders, and particulate flows, where particle shape and thermal effects are of critical importance.

The addition of vibrations during powder application was investigated via DEM simulations with non-spherical, cohesive particles as a new application system. The effect of vibration and process characteristics was explored. Small frequency and amplitude, combined with small recoating velocity, lead to denser granular beds of reduced porosity. Large frequency and amplitude, however, lead to a vibro-fluidized state of the particles and granular beds with loose configuration characterized by an increased porosity. For practical applications, the choice of frequency and amplitude should be considered in combination with a specific translational velocity of the recoating process.




\addcontentsline{toc}{section}{References}
%
%
%
\biblstarthook{In view of the parallel print and (chapter-wise) online publication of your book at \url{www.springerlink.com} it has been decided that -- as a genreral rule --  references should be sorted chapter-wise and placed at the end of the individual chapters. However, upon agreement with your contact at Springer you may list your references in a single seperate chapter at the end of your book. Deactivate the class option \texttt{sectrefs} and the \texttt{thebibliography} environment will be put out as a chapter of its own.\\\indent
References may be \textit{cited} in the text either by number (preferred) or by author/year.\footnote{Make sure that all references from the list are cited in the text. Those not cited should be moved to a separate \textit{Further Reading} section or chapter.} If the citatiion in the text is numbered, the reference list should be arranged in ascending order. If the citation in the text is author/year, the reference list should be \textit{sorted} alphabetically and if there are several works by the same author, the following order should be used:
\begin{enumerate}
\item all works by the author alone, ordered chronologically by year of publication
\item all works by the author with a coauthor, ordered alphabetically by coauthor
\item all works by the author with several coauthors, ordered chronologically by year of publication.
\end{enumerate}
The \textit{styling} of references\footnote{Always use the standard abbreviation of a journal's name according to the ISSN \textit{List of Title Word Abbreviations}, see \url{http://www.issn.org/en/node/344}} depends on the subject of your book:
\begin{itemize}
\item The \textit{two} recommended styles for references in books on \textit{mathematical, physical, statistical and computer sciences} are depicted in ~\cite{science-contrib, science-online, science-mono, science-journal, science-DOI} and ~\cite{phys-online, phys-mono, phys-journal, phys-DOI, phys-contrib}.
\item Examples of the most commonly used reference style in books on \textit{Psychology, Social Sciences} are~\cite{psysoc-mono, psysoc-online,psysoc-journal, psysoc-contrib, psysoc-DOI}.
\item Examples for references in books on \textit{Humanities, Linguistics, Philosophy} are~\cite{humlinphil-journal, humlinphil-contrib, humlinphil-mono, humlinphil-online, humlinphil-DOI}.
\end{itemize}
}
\printbibliography

@article{angelidakis2021clump,
  title={{CLUMP: a Code Library to generate Universal Multi-sphere Particles}},
  author={Angelidakis, Vasileios and Nadimi, Sadegh and Otsubo, Masahide and Utili, Stefano},
  journal={SoftwareX},
  volume={15},
  pages={100735},
  year={2021},
  publisher={Elsevier}
}

@article{castellanos2005relationship,
  title={The relationship between attractive interparticle forces and bulk behaviour in dry and uncharged fine powders},
  author={Castellanos, Antonio},
  journal={Advances in Physics},
  volume={54},
  number={4},
  pages={263--376},
  year={2005},
  publisher={Taylor \& Francis}
}

@article{cundall1979discrete,
  title={A discrete numerical model for granular assemblies},
  author={Cundall, Peter A and Strack, Otto DL},
  journal={Geotechnique},
  volume={29},
  number={1},
  pages={47--65},
  year={1979},
  publisher={Thomas Telford Ltd}
}

@techreport{glasstone1941theory,
  title={The theory of rate processes; the kinetics of chemical reactions, viscosity, diffusion and electrochemical phenomena},
  author={Glasstone, Samuel and Laidler, Keith James and Eyring, Henry},
  year={1941},
  institution={McGraw-Hill Book Company,}
}

@article{gotzinger2003dispersive,
  title={Dispersive forces of particle--surface interactions: direct AFM measurements and modelling},
  author={G{\"o}tzinger, Martin and Peukert, Wolfgang},
  journal={Powder Technology},
  volume={130},
  number={1-3},
  pages={102--109},
  year={2003},
  publisher={Elsevier}
}

@article{gotzinger2004particle,
  title={Particle adhesion force distributions on rough surfaces},
  author={G{\"o}tzinger, Martin and Peukert, Wolfgang},
  journal={Langmuir},
  volume={20},
  number={13},
  pages={5298--5303},
  year={2004},
  publisher={ACS Publications}
}

@article{harrington2018anisotropic,
  title={Anisotropic particles strengthen granular pillars under compression},
  author={Harrington, Matt and Durian, Douglas J},
  journal={Physical Review E},
  volume={97},
  number={1},
  pages={012904},
  year={2018},
  publisher={APS}
}

@article{harrington2020stagnant,
  title={Stagnant zone formation in a {2D} bed of circular and elongated grains under penetration},
  author={Harrington, Matt and Xiao, Hongyi and Durian, Douglas J},
  journal={Granular Matter},
  volume={22},
  pages={1--9},
  year={2020},
  publisher={Springer}
}

@article{jagota1993viscosities,
  title={Viscosities and Sintering Rates of a Two-Dimensional Granular Composite},
  author={Jagota, Anand and Scherer, George W},
  journal={Journal of the American Ceramic Society},
  volume={76},
  number={12},
  pages={3123--3135},
  year={1993},
  publisher={Wiley Online Library}
}

@article{johnson1971surface,
  title={Surface energy and the contact of elastic solids},
  author={Johnson, Kenneth Langstreth and Kendall, Kevin and Roberts, aAD},
  journal={Proceedings of the Royal Society of London. A. Mathematical and Physical Sciences},
  volume={324},
  number={1558},
  pages={301--313},
  year={1971},
  publisher={The Royal Society London}
}

@article{kloss2012models,
  title={Models, algorithms and validation for opensource DEM and CFD--DEM},
  author={Kloss, Christoph and Goniva, Christoph and Hager, Alice and Amberger, Stefan and Pirker, Stefan},
  journal={Progress in Computational Fluid Dynamics, an International Journal},
  volume={12},
  number={2-3},
  pages={140--152},
  year={2012},
  publisher={Inderscience Publishers}
}

@article{li2006london,
  title={London-van der Waals adhesiveness of rough particles},
  author={Li, Q and Rudolph, V and Peukert, W},
  journal={Powder Technology},
  volume={161},
  number={3},
  pages={248--255},
  year={2006},
  publisher={Elsevier}
}

@article{luding2005discrete,
  title={A discrete model for long time sintering},
  author={Luding, Stefan and Manetsberger, Karsten and M{\"u}llers, Johannes},
  journal={Journal of the Mechanics and Physics of Solids},
  volume={53},
  number={2},
  pages={455--491},
  year={2005},
  publisher={Elsevier}
}

@article{muller2011collision,
  title={Collision of viscoelastic spheres: Compact expressions for the coefficient of normal restitution},
  author={M{\"u}ller, Patric and P{\"o}schel, Thorsten},
  journal={Physical Review E},
  volume={84},
  number={2},
  pages={021302},
  year={2011},
  publisher={APS}
}

@incollection{NasatoPoeschelParteli:2016,
	address={Ljubljana},
	author={Schiochet Nasato, Daniel and P{\"o}schel, Thorsten and Parteli, Eric J. R.},
	booktitle={Proceedings of the 6th International Conference on Additive Technologies iCAT 2016, N\"urnberg Nov. 29-30},
	editor={Drstven{\v{s}}ek, Igor and Drummer, Dietmar and Schmidt, Michael},
	pages={260-265},
	publisher={Interesansa},
	title={Effect of vibrations applied to the transport roller in the quality of the powder bed during additive manufacturing},
	year={2016}}

@inproceedings{NasatoHeinlHausottePoeschel:2017,
	address={Barcelona},
	author={Nasato, Daniel S. and Heinl, Martin and Hausotte, Tino and P{\"o}schel, Thorsten},
	booktitle={PARTICLES 2017. Proceedings of the V International Conference on Particle-based Methods. 26-28 September 2017, Hannover, Germany},
	editor={Wriggers, Peter and Bischoff, M. and O\~nate, Eugenio and Owen, D. R. J. and Zohdi, Tarek},
	pages={429-439},
	publisher={CIMNE},
	title={Numerical and experimental study of the powder bed characteristics in the recoated bed of the additive manufacturing process},
	year={2017}}

@article{nasato2020influence,
  title={Influence of particle shape in additive manufacturing: Discrete element simulations of polyamide 11 and polyamide 12},
  author={Nasato, Daniel Schiochet and P{\"o}schel, Thorsten},
  journal={Additive Manufacturing},
  volume={36},
  pages={101421},
  year={2020},
  publisher={Elsevier}
}

@article{nasato2021influence,
  title={Influence of vibrating recoating mechanism for the deposition of powders in additive manufacturing: Discrete element simulations of polyamide 12},
  author={Nasato, Daniel Schiochet and Briesen, Heiko and P{\"o}schel, Thorsten},
  journal={Additive Manufacturing},
  volume={48},
  pages={102248},
  year={2021},
  publisher={Elsevier}
}

@inproceedings{parteli2013using,
  title={Using {LIGGGHTS} for performing {DEM} simulations of particles of complex shapes with the multisphere method,"},
  author={Parteli, EJ},
  booktitle={Proc. In: DEM6-6th International Conference on Discrete Element Methods and Related Techniques, Golden USA},
  year={2013}
}

@inproceedings{parteli2013simulation,
  title={{DEM} simulation of particles of complex shapes using the multisphere method: {A}pplication for additive manufacturing},
  author={Parteli, Eric JR},
  booktitle={{AIP} Conference Proceedings},
  volume={1542},
  pages={185--188},
  year={2013},
  organization={American Institute of Physics}
}

@article{parteli2014attractive,
	author={Parteli, Eric J. R. and Schmidt, Jochen and Bl\"umel, Christina and Wirth, Karl-Ernst and Peukert, Wolfgang and P\"oschel, Thorsten},
	journal={Nature - Scientific Reports},
	pages={6227},
	title={Attractive particle interaction forces and packing density of fine glass powders},
	volume={4},
	year={2014}}

@article{parteli2016particle,
	author={Parteli, Eric J. R. and P{\"o}schel, Thorsten},
	issn={0032-5910},
	journal={Powder Technology},
	pages={96-102},
	title={Particle-based simulation of powder application in additive manufacturing},
	volume={288},
	year={2016}}

@article{parteli2017particle,
	author={Parteli, Eric J. R. and P{\"o}schel, Thorsten},
	journal={European Physical Journal. Web of Conferences},
	pages={15013},
	title={Particle-based simulations of powder coating in additive manufacturing suggest increase in powder bed roughness with coating speed},
	volume=140,
	year=2017}

@article{richard2020predicting,
  title={Predicting plasticity in disordered solids from structural indicators},
  author={Richard, D and Ozawa, Misaki and Patinet, S and Stanifer, E and Shang, B and Ridout, SA and Xu, B and Zhang, G and Morse, PK and Barrat, J-L and others},
  journal={Physical Review Materials},
  volume={4},
  number={11},
  pages={113609},
  year={2020},
  publisher={APS}
}

@article{rieser2016divergence,
  title={Divergence of Voronoi cell anisotropy vector: a threshold-free characterization of local structure in amorphous materials},
  author={Rieser, Jennifer M and Goodrich, Carl P and Liu, Andrea J and Durian, Douglas J},
  journal={Physical Review Letters},
  volume={116},
  number={8},
  pages={088001},
  year={2016},
  publisher={APS}
}

@inproceedings{roy2022local,
  title={Local structural anisotropy in particle simulations of powder spreading in additive manufacturing},
  author={Roy, Sudeshna and Xiao, Hongyi and Shaheen, Mohamad Yousef and P{\"o}schel, Thorsten},
  booktitle={Casablanca International Conference on Additive Manufacturing},
  pages={139--149},
  year={2022},
  organization={Springer}
}

@article{roy2023structural,
	author={Roy, Sudeshna and Xiao, Hongyi and Angelidakis, Vasileios and P\"oschel, Thorsten},
	journal={Additive Manufacturing (submitted)},
	title={Structural fluctuations in thin cohesive particle layers in powder-based additive manufacturing},
	year={2023}}

@article{roy2023effect,
  title={Effect of cohesion on structure of powder layers in additive manufacturing},
  author={Roy, Sudeshna and Shaheen, Mohamad Yousef and P{\"o}schel, Thorsten},
  journal={Granular Matter},
  volume={25},
  number={4},
  pages={68},
  year={2023},
  publisher={Springer}
}

@article{roy2023thermal,
	author={Roy, Sudeshna and Xiao, Hongyi and Angelidakis, Vasileios and P\"oschel, Thorsten},
	journal={(in preparation)},
	pages={},
	title={Thermal modelling of temperature distribution in metal additive manufacturing},
	volume={},
	year={2023}}

@article{schmidt2020packings,
	author={Schmidt, Jochen and Parteli, Eric J. R. and Uhlmann, Norman and W{\"o}rlein, Norbert and Wirth, Karl-Ernst and Peukert, Wolfgang},
	journal={Advanced Powder Technology},
	pages={2293-2304},
	title={Packings of micron-sized spherical particles -- insights from bulk density determination, {X}-ray microtomography and discrete element simulations},
	volume={31},
	year={2020}}

@article{severson2009mechanical,
  title={Mechanical damping using adhesive micro or nano powders},
  author={Severson, BL and Keer, Leon M and Ottino, Julio M and Snurr, Randall Q},
  journal={Powder Technology},
  volume={191},
  number={1-2},
  pages={143--148},
  year={2009},
  publisher={Elsevier}
}

@article{xiao2020strain,
  title={Strain localization and failure of disordered particle rafts with tunable ductility during tensile deformation},
  author={Xiao, Hongyi and Ivancic, Robert JS and Durian, Douglas J},
  journal={Soft Matter},
  volume={16},
  number={35},
  pages={8226--8236},
  year={2020},
  publisher={Royal Society of Chemistry}
}

@article{yu1997modelling,
  title={On the modelling of the packing of fine particles},
  author={Yu, Ai-Bing and Bridgwater, John and Burbidge, A},
  journal={Powder Technology},
  volume={92},
  number={3},
  pages={185--194},
  year={1997},
  publisher={Elsevier}
}

@article{abou2004three,
  title={Three-dimensional particle shape descriptors for computer simulation of non-spherical particulate assemblies},
  author={Abou-Chakra, Hd and Baxter, Jd and T{\"u}z{\"u}n, Ud},
  journal={Advanced Powder Technology},
  volume={15},
  number={1},
  pages={63--77},
  year={2004},
  publisher={Elsevier}
}

@article{kodam2009force,
  title={Force model considerations for glued-sphere discrete element method simulations},
  author={Kodam, Madhusudhan and Bharadwaj, Rahul and Curtis, Jennifer and Hancock, Bruno and Wassgren, Carl},
  journal={Chemical Engineering Science},
  volume={64},
  number={15},
  pages={3466--3475},
  year={2009},
  publisher={Elsevier}
}

@article{cabiscol2018calibration,
  title={Calibration and interpretation of DEM parameters for simulations of cylindrical tablets with multi-sphere approach},
  author={Cabiscol, Ramon and Finke, Jan Henrik and Kwade, Arno},
  journal={Powder Technology},
  volume={327},
  pages={232--245},
  year={2018},
  publisher={Elsevier}
}

@article{marchais20213d,
  title={A 3D DEM simulation to study the influence of material and process parameters on spreading of metallic powder in additive manufacturing},
  author={Marchais, Kevin and Girardot, J{\'e}r{\'e}mie and Metton, Charlotte and Iordanoff, Ivan},
  journal={Computational particle mechanics},
  volume={8},
  number={4},
  pages={943--953},
  year={2021},
  publisher={Springer}
}

@article{haeri2017discrete,
  title={Discrete element simulation and experimental study of powder spreading process in additive manufacturing},
  author={Haeri, Sina and Wang, Y and Ghita, O and Sun, J},
  journal={Powder Technology},
  volume={306},
  pages={45--54},
  year={2017},
  publisher={Elsevier}
}

@article{he2020linking,
  title={Linking particle properties to layer characteristics: Discrete element modelling of cohesive fine powder spreading in additive manufacturing},
  author={He, Yi and Hassanpour, Ali and Bayly, Andrew E},
  journal={Additive Manufacturing},
  volume={36},
  pages={101685},
  year={2020},
  publisher={Elsevier}
}

@article{ostanin2023rigid,
  title={Rigid Clumps in the MercuryDPM Particle Dynamics Code},
  author={Ostanin, Igor and Angelidakis, Vasileios and Plath, Timo and Pourandi, Sahar and Thornton, Anthony and Weinhart, Thomas},
journal={Computer Physics Communications (in press)},
%journal={arXiv preprint arXiv:2310.05027},
year={2023}
}

@inproceedings{laumer2015influence,
  title={Influence of temperature gradients on the part properties for the simultaneous laser beam melting of polymers},
  author={Laumer, Tobias and Stichel, Thomas and Schmidt, Michael},
  booktitle={Proceedings of Laser in Manfacturing Conference 2015, June 22-June 25, 2015 Munich, Germany},
  year={2015}
}

@article{li2021experimental,
  title={Experimental parameter identification for part-scale thermal modeling of selective laser sintering of PA12},
  author={Li, Chao and Snarr, Scott E and Denlinger, Erik R and Irwin, Jeff E and Gouge, Michael F and Michaleris, Pan and Beaman, Joseph J},
  journal={Additive Manufacturing},
  volume={48},
  pages={102362},
  year={2021},
  publisher={Elsevier}
}

@article{peyre2015experimental,
  title={Experimental and numerical analysis of the selective laser sintering (SLS) of PA12 and PEKK semi-crystalline polymers},
  author={Peyre, Patrice and Rouchausse, Yann and Defauchy, Denis and R{\'e}gnier, Gilles},
  journal={Journal of Materials Processing Technology},
  volume={225},
  pages={326--336},
  year={2015},
  publisher={Elsevier}
}

@article{zhao2023multiscale,
  title={Multiscale heat transfer affected by powder characteristics during electron beam powder-bed fusion},
  author={Zhao, Yufan and Aoyagi, Kenta and Cui, Yujie and Yamanaka, Kenta and Chiba, Akihiko},
  journal={Powder Technology},
  volume={421},
  pages={118438},
  year={2023},
  publisher={Elsevier}
}

\backmatter

\include{solutions}
\printindex

\end{document}